# Best Practices for Facing the Security Challenges of Internet of Things Devices Focusing on Software Development Life Cycle

Md Rafid Islam*, Ratun Rahman
Software Engineering, Islamic University of Technology
Gazipur, Bangladesh
Email: rafidislam@iut-dhaka.edu

*Abstract*

In the past few years, the number of IoT devices has grown substantially, and this trend is likely to continue. An increasing amount of effort is being put into developing software for the ever-increasing IoT devices. Every IoT system at its core has software that enables the devices to function efficiently. But security has always been a concern in this age of information and technology. Security for IoT devices is now a top priority due to the growing number of threats. This study introduces best practices for ensuring security in the IoT, with an emphasis on guidelines to be utilized in software development for IoT devices. The objective of the study is to raise awareness of potential threats, emphasizing the secure software development lifecycle. The study will also serve as a point of reference for future developments and provide a solid foundation for securing IoT software and dealing with vulnerabilities.

## 1. Introduction

Online threats and breaches have proved to be a real challenge for securing digital devices. IoT devices include more than just smartphones and computers; they can be almost anything that connects to the internet and communicates [1]. The number and variety of IoT devices are increasing daily, along with the vast amount of data they generate. Now the challenge is to keep this significant amount of data secure because there is always a looming threat of this data being stolen by potential cybercriminals [2]. IoT is used in various sectors, including the medical sector, and a cyberattack on the Internet of Medical Things can expose sensitive data about a patient and jeopardize his safety. By hacking into a smart home or smart car system, the safety of an individual can be compromised as well [3]. That is why ensuring IoT security is of utmost importance.

It is crucial to secure internet-connected devices and networks from cyber threats and intrusions. It can be accomplished by detecting, analyzing, and resolving potential security flaws across all devices [4]. Utilizing Safe Software Development Life Cycle (SSDLC) techniques to develop software in a secure manner is an efficient strategy to defend against threats [5]. The software development lifecycle consists of several stages: planning, design, building, testing, release, maintenance, update, etc. What secure SDLC does is integrate security into all these stages mentioned. By establishing some software development principles and guidelines, several issues and challenges regarding IoT security can be tackled [6].

In section 2, the concept of IoT devices is discussed, followed by security and types of threats to IoT devices. Relative works are mentioned after that. Security aspects in the IoT are divided into 3 categories- people, processes, and technologies, and then broken down into various subcategories in section 3. Sections 4 and 5 cover the result and conclusion.

## 2. Literature Review

This section will focus on introducing the new terms of the research as well as the previous experiments and research that were conducted in a similar field.

**2.1 Background Study:**

2.1.1 IoT devices:

IoT stands for the 'internet of things', which represents physical devices with modern technologies [7]. These devices use the internet or other communication networks to connect and communicate with other devices and exchange information [8]. For example, smart doors use cards to read the data within them as a key and verification method. As a result, people do not have to carry a door key anymore. Some household devices have a common remote control system through a mobile phone. Therefore, a user can simply use that mobile phone to control all other technologies within its range. This reduces the complexity of lifestyles, and a user can control every technology with ease [9]. As the usage of modern devices is increasing rapidly and more and more devices are being added every year, it is safe to say that IoT is the future [10].

2.1.2 Securities for IoT devices

As the IoT may control critical functionalities and use the internet or other communication networks to work, the devices can be extremely vulnerable and a potentially high target for a hacker's attack [11]. Even the data from these devices can be used to exploit the user's lifestyle and other important information. This is why security is a major concern for the IoT. Therefore, almost all IoT devices have some built-in security features [12]. However, if the user is not careful and knowledgeable enough, a hacker can bypass the security. The U.S. Department of Defense has promulgated the Five Pillars of IoT Security, which are: confidentiality, integrity, availability, authenticity, and non-repudiation [13].

2.1.3 Types of threats and attacks for IoT devices

According to byos, six common attacks can be conducted on IoT devices. These attacks include
- Botnets

Botnets are typical malware attacks where an attacker sends them through an email or device and the user accidentally or unknowingly accesses them. There are many anti-malware tools available, especially for botnets. However, they require storage space, which is very difficult to manage on IoT devices [14].
- Ransomware

Ransomware on IoT devices mainly impacts the core functionalities and blocks them from functioning properly [15]. For example, attacking a security camera and stopping it from recording footage.

- Convergence

All IoT devices are connected to some type of network to be controlled remotely [16]. Having a separate network to control the devices could stop coverage; however, people's demand has increased, and they want to control it from anywhere. Thus, the Internet becomes the only viable option. This then provides attackers with an easy option to bypass the internet protocol and control their functionalities if the user makes a human error [17].

- Invisibility

As the usage of the devices is continuously increasing, it is becoming very difficult to monitor all the processes and traffic for abnormal patterns and threats [18].

- Unencrypted data

Most IoT devices do not have an encrypted way to record data to save cost and complexity. They are mostly dependent on some types of clouds, which are particularly vulnerable to eavesdropping, espionage, and hijacking [19].

- Rogue Devices

Rogue devices are solely built on doing harm to a system and stealing data and information. As many people are still unaware of IoT devices and how they perform, it is easier to install an extra or auxiliary device within an IoT device that goes undetected [20].

2.1.4 Software Development Life Cycle

The software development lifecycle (SDLC) is a methodological, multistage process that aims to produce productive and reliable systems in accordance with their functional and design specifications. When figuring out risk, the whole IoT ecosystem needs to be taken into account. This includes both the Internet and the physical systems outside of IoT that use it. So, it is clear that IoT software must be secured in a systematic way throughout the lifecycle of IoT systems and services if they are to provide solutions that are reliable and failsafe. In this regard, it is necessary to secure the SDLC as a whole and take the appropriate factors into account [21].

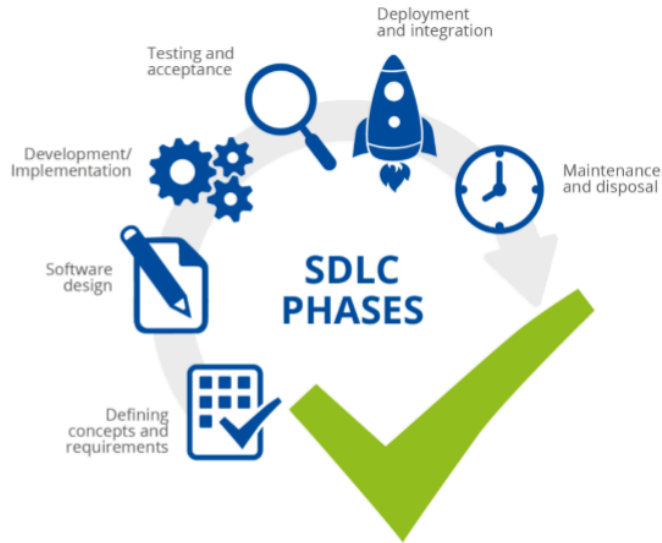

Figure 1: SDLC Phases

**2.2 Relative Work:**
There have been some works and research that introduce and provide information about the security of IoT devices, mostly in recent years. 'Internet of Things' name was first mentioned by Kevin Ashton in 1999 when he proposed to put radio frequency identification chips on products for their tracking in a supply chain [22].

In order to observe who was at his door while he was working in the garage, Jamie Siminoff created the Ring doorbell in 2011, as he kept missing deliveries because he couldn't hear the doorbell from the garage [23].

IoT keeps getting smarter after the introduction of the smart city in 2012. Smart City Switzerland has introduced over 60 projects, including traffic and air quality monitoring, smart transportation and parking, smart buildings, security, heating, lighting, etc. [24].

In 2015, IoT was connected with smartphones, which enabled new features and modifications to the devices as well as adding more devices to IoT. As mobile phones are the most common devices, they helped to connect all other modern devices, and the IoT industry has grown rapidly ever since [25].

In October 2021, May Mobility launched a pilot program to test its self-driving software. Self-driving vehicles are recent additions to the IoT [26].

Internet of Things (IoT): A literature review by Madakam, S., Lake, V has given a brief idea about IoT, its importance, and the future of IoT devices [27]. IoT security: ongoing challenges and research opportunities by Zhang, Z.K., Cho, M.C.Y., Wang, C.W., Hsu, C.W., Chen, C.K. and Shieh, S talked bout the threats and security issues of IoT [28]. Several surveys have also been conducted on the security of IoT. A survey on IoT security: application areas, security threats, and solution architectures by Hassija, V., Chamola, V., Saxena, V., Jain, D., Goyal, P. and Sikdar, B. [29] and The internet of things security: A survey encompassing unexplored areas and new insights by Omolara, A.E., Alabdulatif, A., Abiodun, O.I., Alawida, M., Alabdulatif, A. and Arshad, H. [30] are some examples. Security of IoT application layer protocols: Challenges and findings by Nebbione, G. and Calzarossa, M.C. has also added more information about the attacks on IoT devices and how to prevent them [31].

## 3. Research Methodology

This section of the paper is going to focus on the good practices and guidelines that should be followed to address the threats that have a detrimental impact on the software development lifecycle of the IoT. The most frequently mentioned topics with regard to IoT security are grouped into three main categories, and they are broken down into subcategories, which are discussed in detail below:

**3.1 People:** From the software developers to the end users of the product, "people" are security issues that impact all stakeholders involved in the life cycle of IoT software solutions.

### Training and Awareness

Setting an organizational plan for specialized security training and raising awareness among all the personnel involved is the first step [32]. At all organizational levels, awareness should be increased. After that, an assessment should be made to determine if there is a need for skills or if resources need to be updated to keep them aligned with the latest developments. If the need arises, adequate resource allocation should be ensured by the organization to face any new threats.

### Roles and Privileges

Appropriate people should be assigned after the roles are defined and the minimum level of privilege required for the roles is identified. After that, responsibilities and activities should be separated for the teams. Security controls should be implemented to prevent privilege abuse in the process, and resources need to be allocated to monitor it. A security officer must also be assigned to safeguard the physical facilities.

### Security Culture

Incentives should be given to certain skilled people so that the organization can retain them. Along with that, to improve the process, security experts should be consulted. Threats or risks may arise at any time, necessitating, in addition to monitoring, a prompt response to security incidents.

**3.2 Processes:** When a software project is conceptualized, developed, and brought to market, secure development processes ensure that security concerns are addressed.

**Third-party Management**

Integrity needs to be ensured by setting up a plan for managing the supply chain that includes a security framework. Software dependability should be evaluated while keeping security in mind. Before integrating third-party processes, they must be tested, and the third-party software components should be verified. Data leakages can be prevented by specifying confidentiality clauses.

**Operations Management**

There should be a well-defined plan put in place to deal with incidents and vulnerabilities. Also, a plan needs to be defined for change, patch, and configuration management.

**SDLC Methodology**

It is important to have an authorization policy in place so that only authorized users can use restricted resources. The automation of SDLC processes will decrease human error and effort. Additionally, testing must be incorporated into each phase of the SDLC.

**Secure Deployment**

To dispose of the solution, along with its data and components, a disposal strategy is required. To deal with evolving threats, a process for monitoring and updating should be established. Along with that, an automated testing and secure deployment strategy should be utilized.

**Security Design**

Firstly, a security framework and the least privilege principle are required. The security controls should be verified, and a design review should be conducted. After that, it is necessary to specify security requirements and conduct a risk assessment. The implementation of threat modeling and data classification is also essential.

**Internal Policies**

A communication plan should be established so that the entire organization is aware of the security measures. To prevent information disclosure, it is necessary to implement security controls. It is also important to make sure that the security documents are up-to-date and to have backup plans ready in case some resources aren't available.

**3.3 Technologies:** Technologies are technological tools and procedures used in the software development process to minimize vulnerabilities and defects.

**Access Control**

In addition to storing user credentials securely, it is essential to implement authorization to ensure that applications have the appropriate permissions. Physical protection should be deployed to prevent physical damage.

**Third-Party Software**
When utilizing third-party software components, it is essential that they are patched for the most recent threats discovered. It is better to rely on known secure frameworks.

**Secure Communication**
For secure communication using proven encryption techniques, implementing secure web interfaces and session management is necessary.

**Secure Code**
Secure coding practices and software development techniques should be implemented, and audit capability should be provided in the development phases. Countermeasures are required against rogue code and unauthorized code modification.

**Security Reviews**
It should be ensured that source code is evaluated in terms of security and that attack surface analysis is performed, followed by IoT SDLC tests. In the event that any SDLC phase is interrupted, a contingency plan should be in place.

**Security of SDLC Infrastructure**
Logs should be kept of the different tools used in SDLC phases and IoT systems. A physical detection system and a mitigation plan are required to protect SDLC infrastructure.

**Secure Implementation**
The first step is to modify the default settings and then put restrictions on component customization to prevent losing security functionalities. Additionally, end users should be provided with secure configuration.

## 4. Result and Discussion

It is evident that people, procedures, and technologies all play vital roles in the security measures for IoT devices. Recent years have seen a significant increase in malware and other types of cyberattacks [32]. Both the consumer and the developer of IoT devices will surely benefit from using the security protocols discussed in this article to address some security concerns. To increase knowledge and adjust the development process appropriately, additional research must be done on the specifics of these protocols. It is challenging to switch to a new, modern system nonetheless [33]. But because the Internet of Things (IoT) is so new and has lately become popular, it is imperative to introduce these protocols. A more effective security defense against vulnerabilities can be achieved by using the software development life cycle (SDLC). This approach can also be used to address a number of additional IoT device issues. So, a significant outcome of this study is the introduction and utilization of secure SDLC in IoT devices.

## 5. Conclusion

It is crucial to assess the ever-increasing cybersecurity threats and put in place suitable countermeasures to deal with the typical vulnerabilities that can result from unsecure practices in various SDLC phases. The purpose of the study was to give IoT software developers, maintainers, testers, and system engineers something they could use as a reference and benefit from in the long run. During different stages of software development, such as design, implementation, testing, integration, and maintenance, guidelines and principles to keep in mind for ensuring security are pointed out.